\documentstyle{elsart}
\begin{document}
\begin{frontmatter}
\title{Ab initio molecular dynamics simulations of Aluminum solvation}

\author[jolla]{M.I. Lubin\thanksref{LubinEmail}},
\author[richland]{E.J. Bylaska},
 and
\author[jolla]{J.H. Weare}
\address[jolla]{Department of Chemistry and Biochemistry,
University of California San Diego, La Jolla, CA 92093-0340}
\address[richland]{Environmental Molecular Sciences Laboratary, Pacific
Northwest National Laboratory, P.O. Box 999, Richland, WA 99352}
\thanks[LubinEmail]{E - mail: mlubin@ucsd.edu}

\begin{abstract}

     The solvation of  Al$^{3+}$ and its hydrolyzed species in water
clusters has been studied  by means of {\it ab initio} molecular dynamics
simulations.  The hexa-hydrate Al(H$_{2}$O)$_{6}^{3+}$
ion formed a stable complex in the finite temperature cluster simulation
of one aluminum ion and 16 waters.  The average dipole
moment of strongly polarized hydrated water molecules in the
first solvation shell of Al(H$_{2}$O)$_{6}^{3+}$ was found to be
5.02 Debye.  The deprotonated Al(H$_{2}$O)$_{2}$(OH)$_{4}^{-}$ complex
evolves into a tetra-coordinated Al(OH)$_{4}^{-}$ aluminate
ion with two water molecules in the second solvation shell forming
hydrogen bonds to the hydroxyl groups in agreement with the observed
coordination.  At high temperature and for low water coordination protons
in the first solvation shell are very mobile leading to the formation of
hydrolysis species consistent with the acidity of Al$^{3+}$ in water.

\end{abstract}
\end{frontmatter}

\section{Motivation}

\label{sec:mot}

The aqueous chemistry of aluminum, the most abundant metal in the earth's
crust, remains a subject of fundamental research after many years of
intensive research\cite{may}. Due to the increased acidity of many natural
waters and the corresponding increase in solubility of aluminum containing
minerals, environmental and health issues have further focus attention on
the solution behavior of this toxic material~\cite{martin94,angeletti}. In
addition aluminum hydrolysis products have many practical applications,
ranging from the pharmaceutical design to purification of water\cite{may}.

The aluminum ion, Al$^{3+}$, exists as an octahedral (6-fold coordinated)
hexahydrate ion, Al(H$_{2}$O)$_{6}^{3+}$ in acidic solutions~\cite{marques}.
Over a narrow pH range, $5.5\leq {\rm pH}\leq 6.2$, the hexahydrate
undergoes hydrolysis by four successive deprotonations, producing species
with uniformly decreasing coordination numbers\cite{martin91}. The final
product of the hydrolysis process, the aluminate ion, Al(OH)$_{4}^{-}$, is
tetrahedrally (4-fold) coordinated\cite{martin91}. The coordination behavior
of ions in solution is a sensitive function of the properties of the ions
and the solvating molecules and varies widely from species to species.
For example, the hydrolyzed aqueous ferric ion, Fe$^{3+}$, remains
six-coordinated after four deprotonations~\cite{martin91}.  
The change in coordination of Al(H$_{2}$O)$_{2}$(OH)$_{4}^{-}$
from octahedral to tetrahedral form occurs to
accommodate changes in ligands charges and polarizations~\cite{martin91}.
Because of the small relative size of the Al$^{3+}$ ion,
this change in
coordination both strongly polarizes the OH bond of water ligands and
stabilizes the formation of OH$^{-}$. Both these effects should increase
acidity constant of Al$^{3+}$ in water solutions, which is about 10$^{-5}$
at room temperature~\cite{baes}. However, the microscopic description of
these processes, especially the mechanisms of hydrolysis, remains incomplete.

We will show in this Letter using {\it ab initio} molecular dynamics
 (AIMD)~\cite{car-parrinello}
 that this strong coordination interaction results in
highly inhomogeneous electron charge distribution in the first solvation
shell of solvating water molecules.
 D. Marx {\it et al.} have
 successfully applied a similar method to the studies of the
structural and dynamical properties of bivalent aqueous Be$^{2+}$ ion~\cite
{marx}. 
However, there have been a number of efforts to simulate highly charged
ions in solution using traditional molecular dynamics methods.
Given the complexity of many-body interactions in such systems with
highly charged ions and hydrogen bonded water molecules, the development of
phenomenological potentials describing the interaction of trivalent ions 
(Al$^{3+}$, Cr$^{3+}$) with water and further classical molecular dynamics
simulations often neglect hydrolysis~\cite{wasserman,martinez}. If the goal
is to model the solvation of Al$^{3+}$ in the practically important range of
pH $=4-7$, hydrolysis has to be taken into account to correctly describe the
stability of the surrounding aluminum water network~\cite{eigen61,eigen63}.

\section{Technical details}

\label{sec:tech}

In this Letter we illustrate the essential features of Al$^{3+}$ hydrolysis
in terms of the hydration numbers and structural parameters of the species
in aqueous solutions, namely, Al(H$_{2}$O)$_{6}^{3+}$ and Al(OH)$_{4}^{-}$.
  We will show that the efficient local density approximation (LDA),
which has been used throughout the paper,
provides accurate structural information for
both species (see Section~\ref{sec:res}).  
Vosko {\it et al}'s parameterized form for the
exchange-correlation energy has been used to implement LDA~\cite{vosko}.
Since the potential energy surfaces of small clusters usually have many
local minima, we have used the method of dynamical simulated annealing to
find the optimal structures on the ground state potential surface. Such a
search for a global minima would be difficult to do using only conventional
total energy methods such as steepest descent minimization. To get accurate
energetics, the Perdew-Burke-Ernzerhof 1996 (PBE96) generalized gradient approximation~\cite{pbe} has been employed in a postprocessing mode.

The valence electronic wave functions were expanded in plane waves, and
their interactions with nuclei and the core electrons were described through
generalized norm-conserving Hammann pseudopotentials~\cite{hammann}. The
nonlocal part of the pseudopotentials was modified to a completely separable
form as suggested by Kleinman and Bylander~\cite{kleinman}. Since the
original Hammann pseudopotential requires too high cut-off energy for
oxygen, a softer potential was constructed by increasing the core radii.
Using this potential, the equilibrium bond distances and the binding
energies of single water molecule and the water dimer were tested against
known LDA results~\cite{salahub92} at the cutoff energy ${\rm E_{cut} = 100}$
Ryd (see Table~\ref{tab:test}). 
\begin{table}[p]
\caption{Intramolecular structural parameters for water monomer and hydrogen
bond parameters for water dimer in comparison with known LDA results and
experimental data (in \AA \ and deg).}
\label{tab:test}
\vspace*{0.5in}
\par
\begin{center}
\begin{tabular}{lccc}
\hline
& calculated results \hspace*{0.15in} & Ref.~\cite{salahub92}
\hspace*{0.15in} & experiment \hspace*{0.15in} \\ \hline
water monomer &  &  &  \\ 
r$_{{\rm OH}}$ & 0.953 & 0.978 & 0.957$^{a}$ \\ 
$< {\rm HOH}$ & 104.8 & 104.4 & 104.5$^{a}$ \\ \hline
water dimer &  &  &  \\ 
r$_{{\rm OO}}$ & 2.713 & 2.710 & 2.98 $^{b}$ \\ 
$< {\rm O}-{\rm H_{\ldots}O}$ & 170.8 & 171 & 174 $\pm 10$ $^{b}$ \\ \hline
\end{tabular}
{\\[0pt]
{\small $^{a)}$ From Ref.~\cite{benedict56}. \hspace*{0.2in} $^{b)}$ From
Ref.~\cite{odutola80}.}}
\end{center}
\end{table}
Good agreements in these tests indicated that our soft pseudopotential for
the oxygen is accurate. This choice of the cut-off energy required about
130,000 plane waves for each molecular orbital in the 20 a.u cubic cell. 
 We performed our calculations without imposing periodic boundary conditions.
Because we are dealing with charged isolated clusters with the strong
dipoles and long-range Coulomb interactions, an aperiodic convolution method
for solving Poisson's equations for free-space boundary conditions has been
used~\cite{bylaska}. We have also replaced hydrogen atoms by deuterium in
our simulations in order to be able to use sufficiently large values for the
time step (${\Delta t}=$ 7 a.u) and fictitious mass
(${\mathrm \mu =1100}$ a.u.) in Car - Parrinello dynamics.

\section{Results and discussion}

\label{sec:res}

The acidity of a hydrated metal ion depends on the strength of the
coordination bond between cation and the oxygen atoms of the first hydration
shell water molecules. If this coordination bond is strong, the OH bond in
the solvating water molecules of the first coordination shell are strongly
polarized. These polarized bond  can dissociate in a polar solvent to form
hydrolyzed species such as Al(H$_{2}$O)$_{5}$(OH)$^{2+}$. This process is
important to many aspects of aluminum solvation and is complicated by the
many interactions that determine the structure of the coordination shell. In
the following, we will show that the polarization of the solvating molecules
is a function of the coordination number of the first solvation shell. This
leads to a cooperative relation between mobility of the protons in the 
first solvation shell and the coordination structure.

\subsection{{\rm Structure of aqueous Al$^{3+}$}}

\label{subsec:al.struct}

The formation of a stable coordination cell around Al$^{3+}$ occurs on the
subpicosecond time scale and hence well within the range of an {\it ab initio%
} molecular dynamics (AIMD) run. Our AIMD runs were initiated by placing one
aluminum ion in the middle of the cluster of 16 randomly oriented water
molecules in a cubic simulation box with a length of 20 a.u. On the time
scale of about 0.25 ps, dramatic rearrangement of water cluster took place
with the formation of an almost perfect octahedra complex of 
Al(H$_{2}$O)$_{6}^{3+}$
surrounded by the 10 remaining water molecules. This octahedral
structure appears to be very stable and remained intact during the 1 ps run
at ambient conditions. During this run no exchange of water molecules
between first and second solvation shell was observed.

To check the structural parameters of the hexa-hydrate complex as calculated
by our method, we performed AIMD simulations on isolated 
Al(H$_{2}$O)$_{6}^{3+}$
(by removing 10 remaining waters from the structure obtained in
our first run) and also small clusters of [Al(H$_{2}$O)$_{n}$]$^{3+}$ with
 ${\mathrm n=1,2,3,4,5}$. 
Since there are several possible low energy isomers of
these clusters we have used simulating annealing, i.e., heating 
and cooling the system by reducing the
kinetic energy, together with the steepest
descent geometry optimizations to ensure that the lowest energy structures
were obtained. In the present calculation we started from an initial
temperature of 1000$^{\circ}$ K
 in the simulated annealing procedure. By performing
short (on the time scale $\tau =10$ fs) molecular dynamics runs followed by
periodic quenching of the system, a rather fast effective cooling schedule
was enforced. The geometrical parameters obtained in our calculations
(see Table~\ref{tab:aq.al}) are in good agreement with the second order
M{\o}ller-Plesset perturbation theory (MP2) results~\cite{wasserman}.

For ${\rm n=1, 2, 3}$ planar structures were obtained. The ${\rm n=4}$
cluster was a tetrahedral complex. At ${\rm n=5}$, the structure of
Al(H$_{2} $O)$_{5}^{3+}$
was a square-based pyramid (aluminum sits slightly above
the center of the square formed by four waters with fifth water molecule
along the line perpendicular to the square). 
This is different from Ref.~\cite{wasserman},
 where Al(H$_{2}$O)$_{5}^{3+}$ was found to be a stable trigonal
bipyramid (which can be viewed as distorted square-based pyramid) under
 $C_{2v}$ constrained optimization. We tried to find a $C_{2v}$ bipyramid
structure but the system always went to the square pyramid on annealing. The
minimum energy geometries of the trigonal bipyramid Al(H$_{2}$O)$_{5}^{3+}$
and octahedral complex Al(H$_{2}$O)$_{6}^{3+}$ are given in
the Fig.~\ref{fig:struct}. 

Filling the first solvation shell of the aluminum cation by water molecules
leads to gradual increase in the length of the coordination bond Al-O. At
the same time the OH bond length of the waters in the first solvation shell
approaches the value for free H$_{2}$O at ${\rm n = 6}$.
 The intramolecular angle $< {\mathrm HOH}$
for the molecules in the first solvation shell is larger then in a
free water molecule. 
As a function of coordination number the $< {\rm HOH}$ first decreases
and then increases with the minimum at ${\rm n = 3}$. 
This trend is also observed in MP2 data but the numbers are
smaller~\cite{wasserman}.

The cohesive energies per coordination bond for Al(H$_{2}$O)$_{n}^{3+}$
 (${\mathrm n = 1-6}$) clusters have been calculated according to the formula 
\begin{equation}
\Delta E_{\mathrm{n}} =(E(\mathrm{Al}^{3+}(\mathrm{H}_{2}\mathrm{O)}_{\mathrm{n}}-E(%
\mathrm{Al}^{3+})-\mathrm{n}E(\mathrm{H}_{2}\mathrm{O))/n}
\end{equation}
using PBE96 functional (PBE96 results differed by a few percent from the magnitude of the LDA results). The cohesive energies per coordination bond 
$\Delta E_{\mathrm{n}}$ as well as total cohesive energies
$\Delta E_{\mathrm{n}} \times \mathrm{n}$ for each of these clusters are reported in the Table 2 and are in close agreement with
MP2 results~\cite{wasserman}. 
\begin{table}[p]
\caption{Global minimum geometries (in \AA  \ and deg) and average dipole
moments of water molecules (in Debye) of [Al(H$_{2}$O)$_{{\rm n}}$]$^{3+}$
complexes with ${\mathrm n=1, 2, 3, 4, 5, 6}$, obtained with the
LDA density functional as described in the text.
Total  and per coordination bond cohesive energies  (in kcal/mol),
$\Delta E_{\mathrm{n}} \times \mathrm{n}$ and $\Delta E_{\mathrm{n}}$ 
respectively, were calculated using PBE96 functional.
 For comparison, all electron MP2/cc-pwCVTZ results for
${\mathrm n=1, 6}$ and MP2/cc-pVDZ results for
${\mathrm n = 5}$ from Ref.\protect\cite{wasserman} are
reported in parentheses. For the ${\mathrm n = 6}$ the experimental
X-ray diffraction result of Ref.\protect\cite{marques} is given.}
\label{tab:aq.al}
\vspace*{0.5in}
\par
\begin{center}
\begin{tabular}{ccccccc}
\hline
& n = 1 & n = 2 & n = 3 & n = 4 & n = 5 & n = 6 \\ \hline
r$_{{\rm AlO}}$ & 1.724 & 1.751 & 1.751 & 1.815 & 1.823 \hspace{2mm} 1.870 
\hspace{2mm} 1.880 & 1.920 \\ 
& (1.723) &  &  &  & (1.895) (1.901) (1.949) & (1.911) \\ 
&  &  &  &  &  & 1.89 (experiment) \\ \hline
r$_{{\rm OH}}$ & 1.031 & 1.011 & 0.999 & 0.989 & 0.980 & 0.976 \\ 
& (1.018) &  &  &  & (0.984) & (0.972) \\ 
$< {\rm HOH}$ & 106.9 & 106.5 & 106.2 & 106.7 & 107.1 & 107.6 \\ 
& (106.54) &  &  &  & (106.01) & (106.56) \\ \hline
{\rm dipole} & 5.9 & 5.51 & 5.26 & 5.06 & 5.02 & 5.02 \\ \hline
$\Delta E_{\mathrm{n}}$ & -203.7 & -184.3 & -166.45 & -148.16 & -131.25 & 
-118.36 \\ \hline
$\Delta E_{\mathrm{n}} \times \mathrm{n}$ & -203.7 & -368.6 & -499.35 & -592.64
& -656.27 & -710.16 \\ 
& (-201.3) &  &  &  & (-664.2) & (-723.7) \\ \hline
\end{tabular}
\end{center}
\end{table}

To characterize the strength of polarization of the hydrated water
molecules, in addition to the geometrical parameters of the optimized
structures of [Al(H$_{2}$O)$_{n}$]$^{3+}$
 we also determined the approximate dipole moments of
water molecules in these clusters. We found the centers
 $\langle {\mathrm r}_{{\mathrm e}}\rangle$
 of the electronic density for the particular water
molecule according to the formulae 
\begin{equation}
\langle {\mathrm r}_{{\mathrm e}}
\rangle =\frac{\int d\vec{{\mathrm r}}\ \vec{{\mathrm r}}\
\rho (\vec{{\mathrm r}})}{\int d\vec{{\mathrm r}}\ \rho 
(\vec{{\mathrm r}})}
\label{eq:dipole}
\end{equation}
where the integration is over a sphere of about 3 a.u. centered on the
oxygen ions. We note that the electron density was well localized on the
solvating waters. Using this procedure the dipole moment of isolated water
molecule was found to be 1.838 D in good agreement with the experimental
value of 1.855 D~\cite{shepard}. The average dipoles of the single hydrated
waters in [Al(H$_{2}$O)$_{{\rm n}}$]$^{3+}$ clusters with
 ${\mathrm n=1,2,3,4,5,6}$ decrease by almost
 $20\%$ from 5.9 D to 5.02 D as a function of
coordination. These numbers represent a substantial increase from the dipole
moments of water molecule in vacuum and in the bulk liquid which are
 1.855 D~\cite{shepard} and 2.6 D~\cite{coulson}, respectively.

The increase in the bond polarization as a function of hydration numbers
plays an important role in the dynamical behavior of the systems as will be
discussed below. To illustrate the reduction of energy in the transfer of a
proton from water in the first solvation shell to vicinity of 
an oxygen in the second
solvation shell in the presence of aqueous Al$^{3+}$ we carried out the
following calculation. In the low energy structure of
 Al(H$_{2}$O)$_{7}^{3+}$, a proton was moved from the water in the
 first solvation shell to the
nearest water molecule in the second solvation shell along hypothetical
reaction path leading to the acid reaction 
\begin{equation}
\mathrm{Al}^{3+}\mathrm{(H}_{2}\mathrm{O)}_{6}\mathrm{+H}_{2}
\mathrm{O(2nd \ solvation \ shell)}\longrightarrow \mathrm{Al}^{3+}\mathrm{(H}_{2}\mathrm{O)}_{5}\mathrm{OH}^{-}
\mathrm{+H}_{3}\mathrm{O}^{+}.
\label{eq:acid}
\end{equation}
The procedure was done at the constant O-O separation of 2.403 \AA.
We kept O-H-O atoms collinear while moving the proton between 
oxygen atoms in steps of 0.02 \AA. 
The potential energy calculated using the PBE96 functional along
the path O-H-O is plotted in Fig.~\ref{fig:drag} as a function
of the asymmetric stretch coordinate 
\begin{equation}
Q=\frac{|r_{\mathrm{O}_{1}\mathrm{H}}|-|r_{\mathrm{O}_{2}
\mathrm{H}}|}{\sqrt{2}}
\label{eq:strech}
\end{equation}
where $r_{\mathrm{O}_{1}\mathrm{H}}$ and $r_{\mathrm{O}_{1}\mathrm{H}}$
are the
distances between the transferring proton and oxygens of water molecules.
The potential energy for the proton transfer process from the
aqueous Al$^{3+}$ to the near vicinity of the water molecule in
the second solvation
shell is represented by the broad potential well of 9 kcal/mol. We note
that there is only a single minimum in the potential given in Fig.~\ref
{fig:drag}. To form a minimum corresponding to H$_{3}$O$^{3+}$ species may
require a more complex reaction coordinate~\cite{cheng}. This will be the
subject of further research. A similar calculation for proton transfer in a
water dimer was carried out. In this case the transfer of the proton from
one water molecule to the other requires about 50 kcal/mol. This dramatic
reduction in the potential energy landscape due to the presence of aqueous
 Al$^{3+}$ contributes to high mobility of the protons in our molecular
dynamics simulations of aluminum ion with ${\mathrm n=14-16}$ water molecules
discussed below (see Section~\ref{subsec:beyond}).

\subsection{{\rm Structure of aluminate ion, Al(OH)$_{4}^{-}$ }}

\label{subsec:aluminate}

Since the LDA geometry of aqueous aluminum ion, Al(H$_{2}$O)$_{6}^{3+}$
compared well with MP2 data and experimental X-ray diffraction results, we
initiated a study of the formation of the most important hydrolysis product,
Al(OH)$_{4}^{-}$. To do this we started an AIMD run with the equilibrated
Al(H$_{2}$O)$_{6}^{3+}$ structure but with 4 protons removed. The resulting
deprotonated hexa-hydrate complex, Al(OH)$_{4}^{-}$(H$_{2}$O)$_{2}$, 
after 0.1 ps readily evolved towards the equilibrium 
tetrahedral structure of aluminate ion, Al(OH)$_{4}^{-}$,
see Fig.~\ref{fig:aluminate}.
The two remaining water molecules were forced out
of the first solvation shell and formed hydrogen bonds to the hydroxyl
groups coordinated to the aluminate anion. 

Since these simulations were done at the finite ionic temperature, 
it allowed us to test the thermal stability of this process. 
The resulting negatively
charged fourfold coordinated structure has a negative highest occupied
eigenvalue of Kohn-Sham orbital of $\varepsilon = -2.07$ eV. The tetrahedral
structure of Al(OH)$_{4}^{-}$ remained stable with no exchange
 in the following 0.5 ps of the AIMD run. 
 Extensive MP2/6-311G ** and B3LYP/6-311G **
calculations of Al(OH)$_{4}^{-}$ gave the average Al-O distance as 
1.792 \AA~\cite{teppen} which compares well with our LDA result, 1.750 \AA.
The Al-O
bonds are strengthened in tetrahedral Al(OH)$_{4}^{-}$ structure as compared
to the  Al(H$_{2}$O)$_{6}^{3+}$ octahedral complex, where the Al-O distance
is $9\%$ larger. As was discussed in Section~\ref{sec:mot}, the change in
coordination from six in Al(H$_{2}$O)$_{6}^{3+}$ to four in Al(OH)$_{4}^{-}$
is a unique feature of aluminum hydrolysis.

\subsection{\rm Beyond first solvation shell}

\label{subsec:beyond}

The interaction between first and second solvation shells is an important
issue in theoretical~\cite{marx,wasserman} and experimental studies~\cite
{caminiti,beyer} of the solvation of highly charged ions in water. As a
first step to the understanding of these effects in bulk water, we report
here results for the solvation of aluminum ion in the clusters of
 ${\mathrm n = 14 - 16}$ waters.

The presence of hydrolysis (e.g. acid reaction of the form of
Eq.~\ref{eq:acid}) in the Al$^{3+}$ water system is
well documented~\cite{martin94}.  
However, our low temperature simulations with Al$^{3+}$ in
the middle of cluster of water molecules did not show any proton mobility on
the time scale of the simulations. Since the hydrolysis constants in the
aqueous aluminum ion solutions are known to be four orders of magnitude
higher at the temperature of 800$^{\circ}$ K
 then at room temperature~\cite{wesolowski},
 we also performed simulations with the Al$^{3+}$ inside cluster of 14
water molecules in this elevated temperature regime. Hydrolysis did occur in
this heated system on the subpicosecond time scale. 
As the coordination bond of one of the six hydrated
waters (with purple colored oxygen) 
lengthened two protons (green) on the remaining waters of
the first solvation shell shuttled between hydrated
waters and the incomplete second solvation shell. Snapshots showing the
shuttling protons coordinated to second solvation shell water molecules
are presented in  Fig.~\ref{fig:shuttle}. 

In our calculations of the polarization of the OH bonds in the first
coordination shell we have shown that the decrease in repulsive interactions
in the solvation shell for low coordination led to shorter Al-O bond lengths
and corresponding increase in polarization of the OH bonds. In a polarizable
medium (e.g. water)
 this increase in polarization should lead to increased mobility of
the protons. Evidence for this from our simulations is given in
Fig.~\ref{fig:surface} where we show a cluster structure in which
the Al$^{3+}$ ion is coordinated to only 3 waters. The extensive
migration of the protons even at room temperature shown in
this figure support the above conjecture.  
Neighboring waters of Al$^{3+}$
lose their protons shortly ($\sim$~50 fs) after the cation 
has been artificially
moved into the 3-water coordination structure on the surface
of the water cluster.
After equilibration at room temperature, one proton was found to shuttle
between hydrated water and the water in the second solvation shell, while
two other protons left first solvation shell and became engaged in the
hydrogen bonding network of the solvent. 
These three protons which showed high mobility are colored green
in Fig.~\ref {fig:surface}.
Although this ''surface'' structure
of aqueous aluminum is an artificial construction, it suggests the crucial
role of the coordination of Al$^{3+}$ in hydrolysis. 

\section{Acknowledgments}
\label{sec:acknowledgments}

We would like to thank Dr. M. Marron (ONR) for providing us with support and
computer time at Naval Research Center through the ONR project
F49620-94-1-0286. Calculations were performed at the NAVOOCEANO
Supercomputer Center.

\clearpage
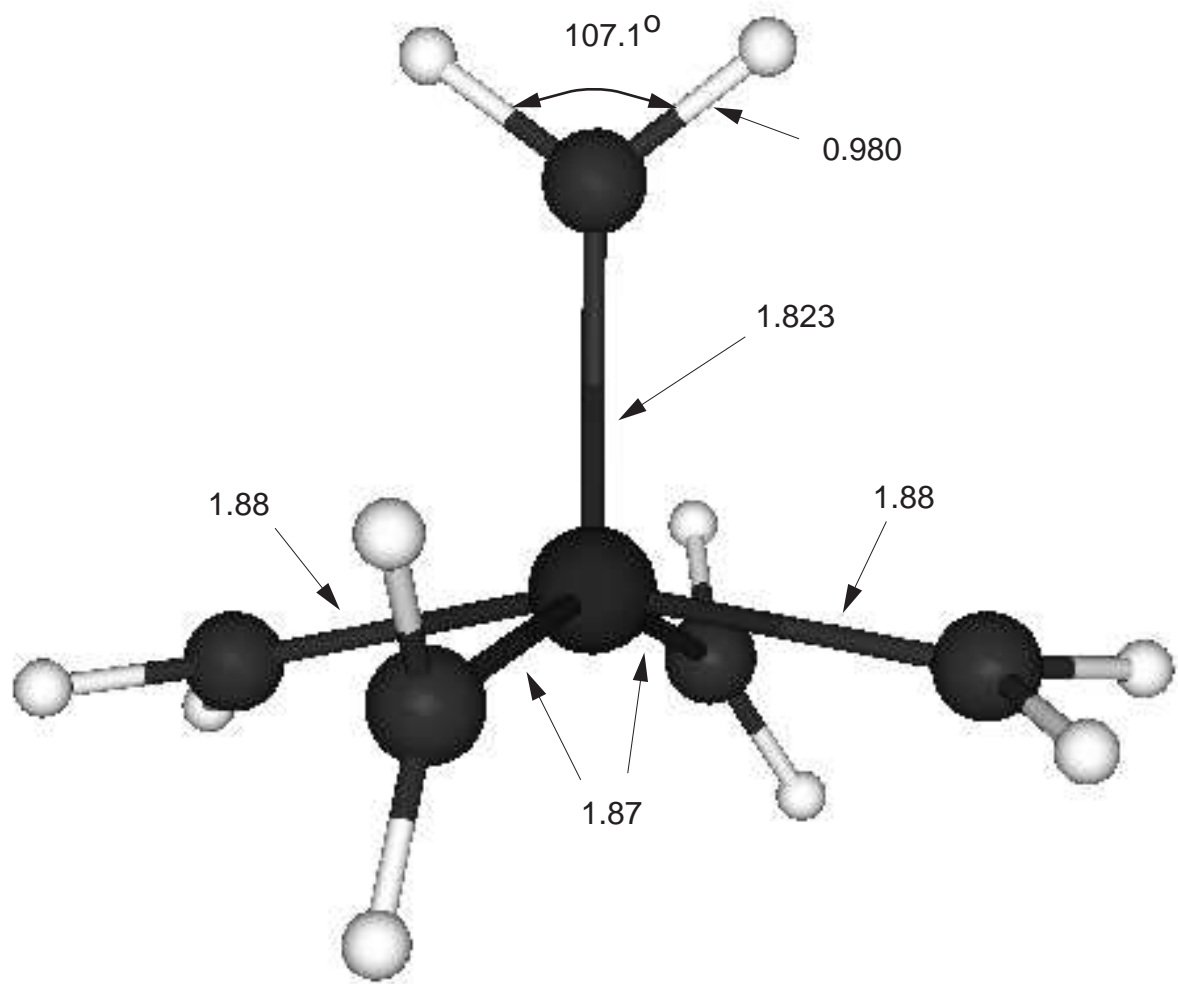
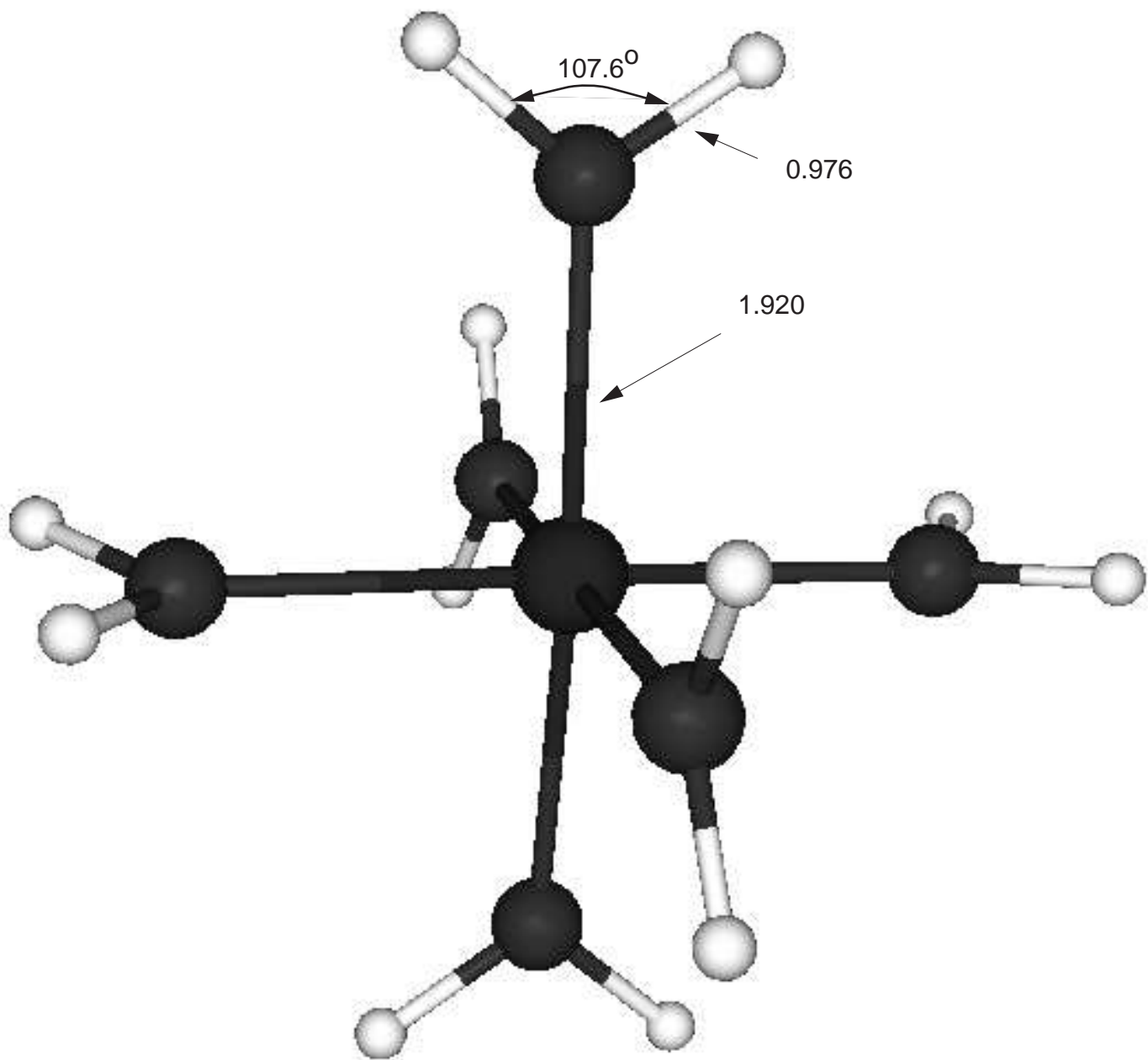
\begin{figure}[tbp]
\caption{The minimum energy geometries of a) Al(H$_{2}$O)$_{5}^{3+}$
and b) Al(H$_{2}$O)$_{6}^{3+}$}~.
\label{fig:struct}
\end{figure}
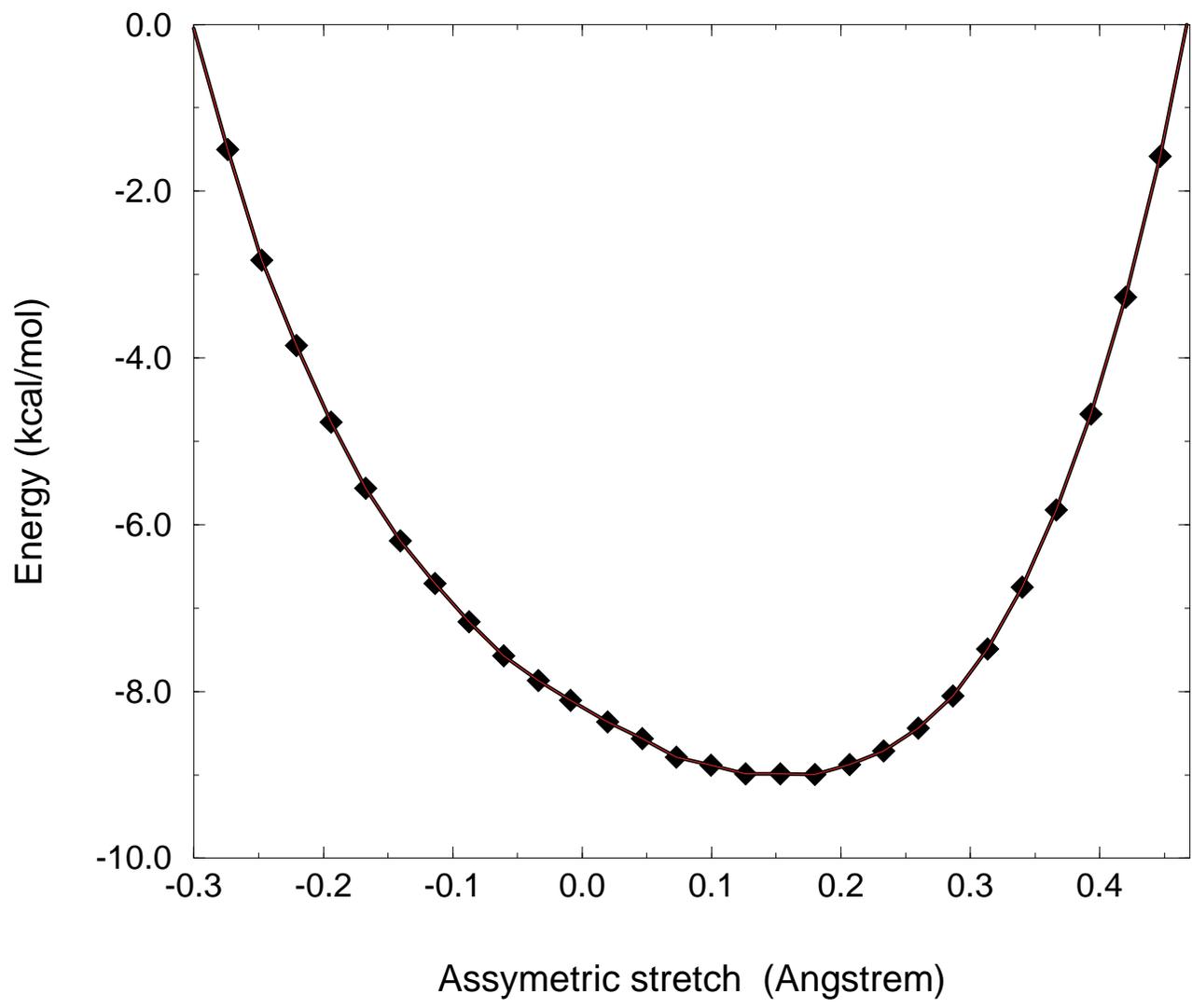
\begin{figure}[tbp]
\caption{Potential energy (in kcal/mol) for the proton moved between two
water molecules in steps of 0.02 \AA \ along a linear path O-H-O in
Al(H$_{2}$O)$_{7}^{3+}$ cluster.  
O-O separation is 2.403 \AA.  
O-H-O atoms have been kept collinear. Q is the asymmetric stretch 
coordinate (see text).}
\label{fig:drag}\vspace{30mm}
\end{figure}
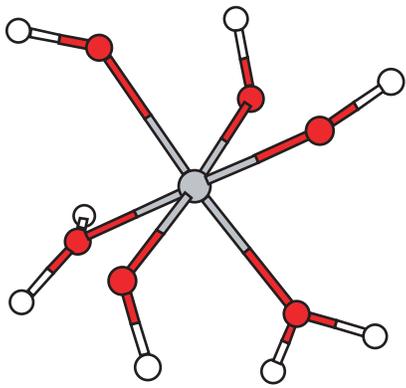
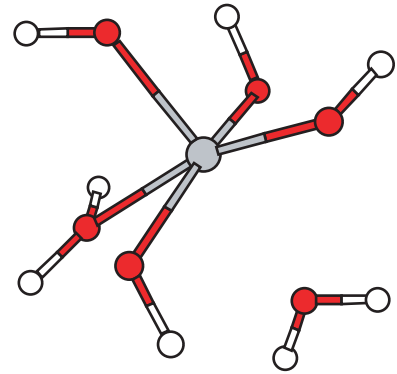
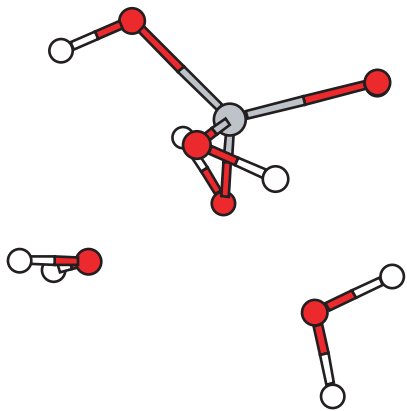
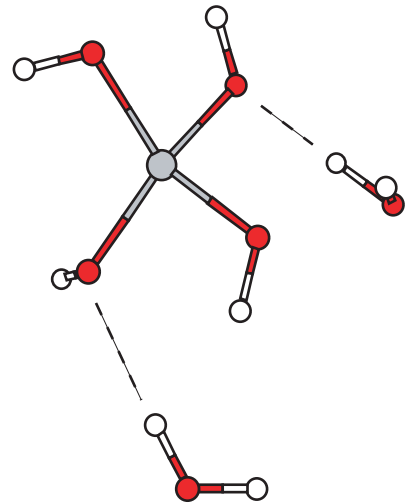
\begin{figure}[tbp]
\caption{Snapshots from {\it ab initio} molecular dynamics simulation
starting from the deprotonated Al(OH)$_{4}^{-}$(H$_{2}$O)$_{2}$ cluster. 
(a) Octahedral initial configuration;
 (b) after 0.025 ps;
(c) after 0.05 ps; and (d) the final tetrahedral structure of Al(OH)$_{4}^{-}$
at 0.3 ps with two remaining water molecules forced out of the
first solvation shell.  Hydrogen bonds are illustrated by 
dashed lines.}
\label{fig:aluminate}\vspace{30mm}
\end{figure}
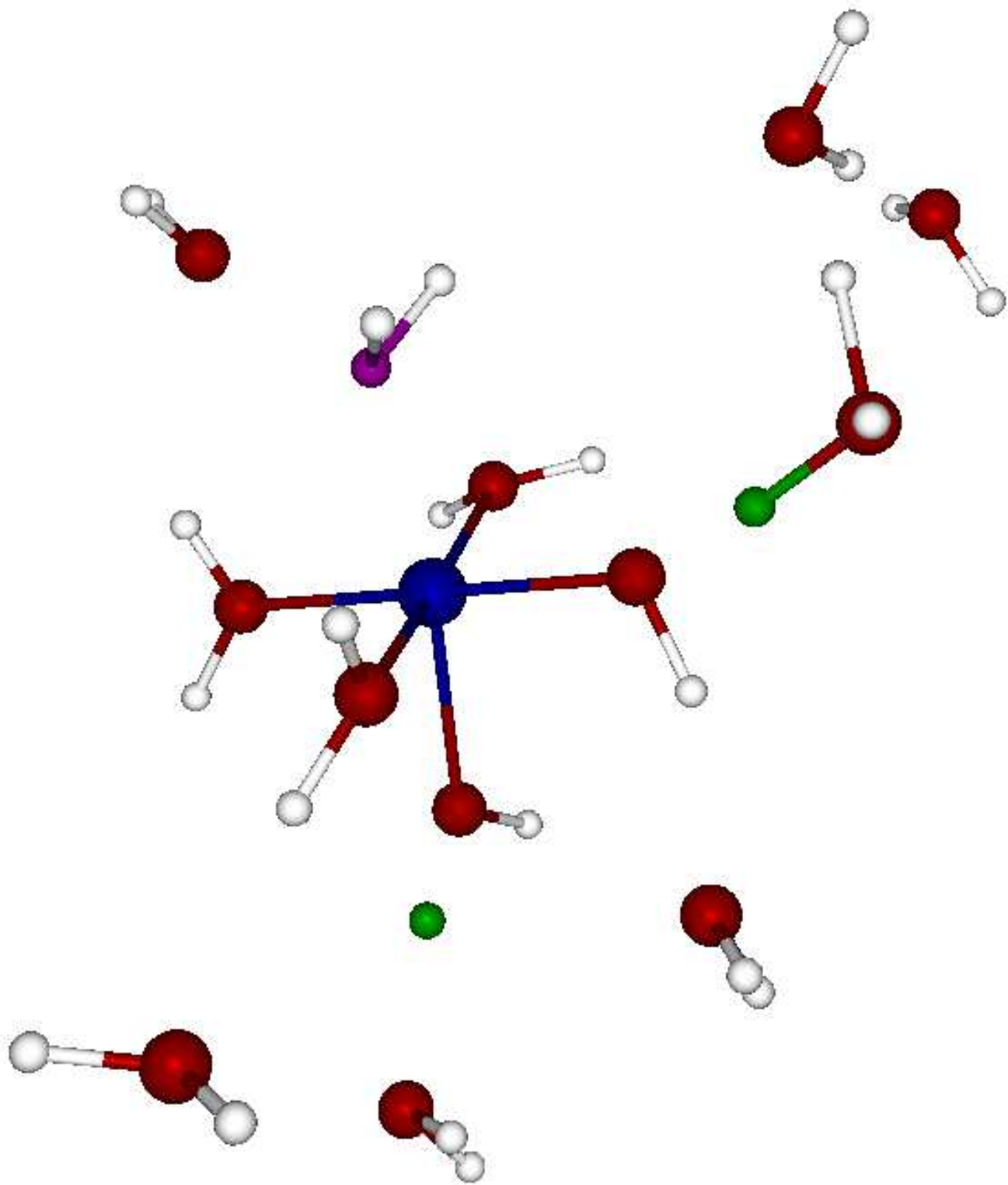
\begin{figure}[tbp]
\caption{Snapshot of the hydrated Al(H$_{2}$O)$_{6}^{3+}$ at a
temperature of about 800$^{\circ}$ K with the protons (green)
shuttling between hydrated waters.}
\label{fig:shuttle}\vspace{30mm}
\end{figure}
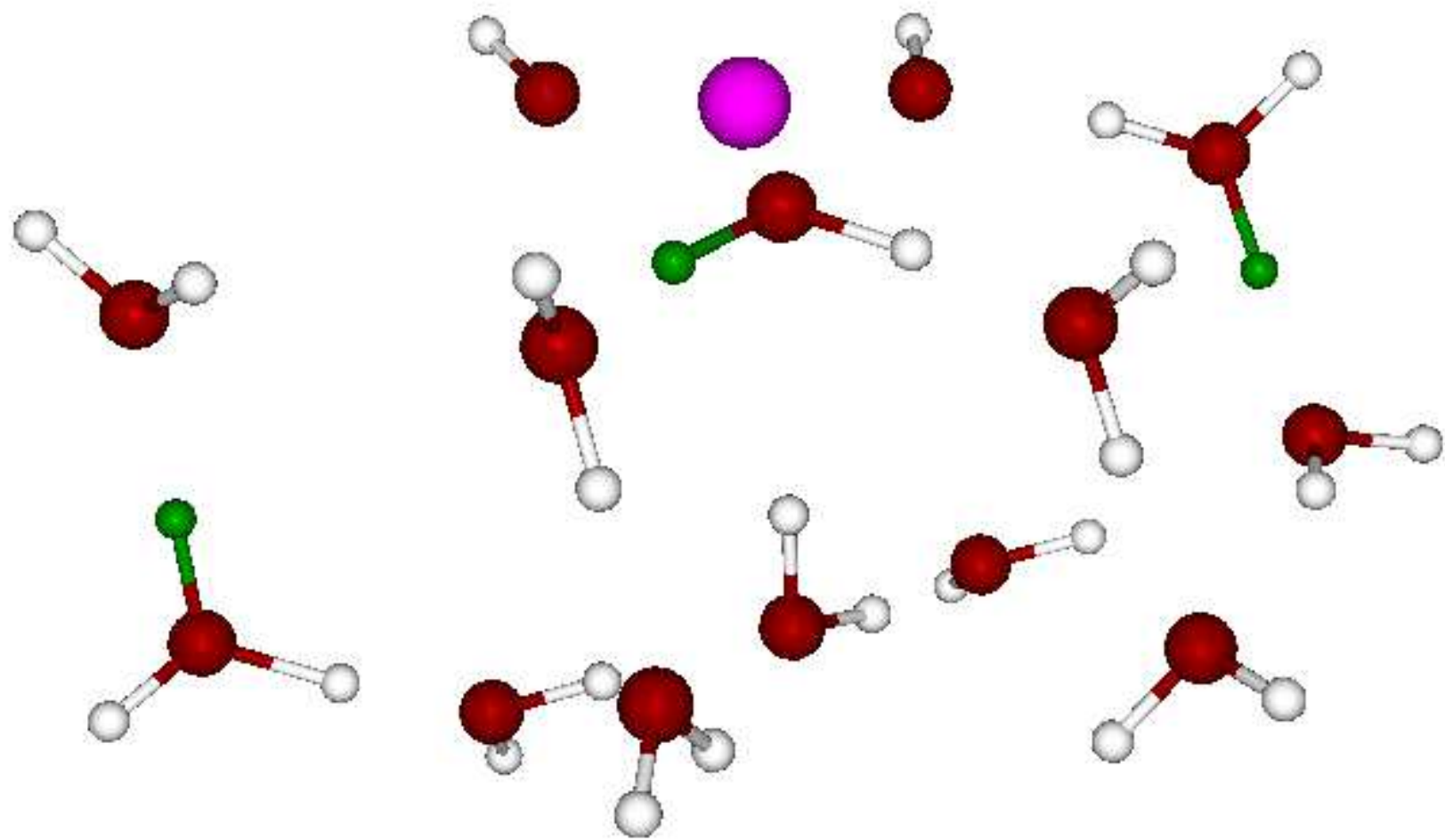
\begin{figure}[tbp]
\caption{Snapshot of the AIMD simulation of the artificial "surface"
structure with a low coordinated aluminum ion on the surface of the water
cluster and extensive hydrolysis.  
Protons which showed high mobility are colored green.}
\label{fig:surface}
\end{figure}

\end{document}